\documentclass[sigplan, 10pt]{acmart}
\renewcommand\footnotetextcopyrightpermission[1]{} 
\usepackage[utf8]{inputenc}
\usepackage[toc,page]{appendix}
\usepackage{tabularx}
\usepackage{xurl}
\usepackage{amsmath}
\usepackage{natbib}

\makeatletter
\renewcommand\@formatdoi[1]{\ignorespaces}
\makeatother

\setcopyright{none}

\renewenvironment{quote}{%
  \list{}{%
    \leftmargin0.75cm   
    \rightmargin\leftmargin
  }
  \item\relax
}
{\endlist}

\newcommand{\blockquote}[2]{
    \blockquotem{\textit{#1}}{#2}
}

\newcommand{\blockquotem}[2]{
\begin{quote}
    {#1}
\end{quote}%
\begin{flushright}- #2.\end{flushright}
}

\setcitestyle{citesep={;}}

\newtheorem*{motivation}{Motivation}

\begin{document}

\title{Extensibility in Programming Languages: An overview}
\author{Sebastian Mateos Nicolajsen}
\affiliation{\institution{IT University of Copenhagen (ITU)}\country{Denmark}}
\date{February 2022}


\begin{abstract}
I here conduct an exploration of programming language extensibility, making an argument for an often overlooked component of conventional language design. Now, this is not a technical detailing of these components, rather, I attempt to provide an overview as I myself have lacked during my time investigating programming languages. Thus, read this as an introduction to the magical world of extensibility. Through a literature review, I identify key extensibility themes - Macros, Modules, Types, and Reflection - highlighting diverse strategies for fostering extensibility. The analysis extends to cross-theme properties such as Parametricism and First-class citizen behaviour, introducing layers of complexity by highlighting the importance of customizability and flexibility in programming language constructs.

By outlining these facets of existing programming languages and research, I aim to inspire future language designers to assess and consider the extensibility of their creations critically.
\end{abstract}

\maketitle
\pagestyle{plain}

\section{Introduction}

\blockquote{Pascal is for building pyramids—imposing, breathtaking, static structures built by armies pushing heavy blocks into place. Lisp is for building organisms.}{Alan Perlis \cite{gab2012}}

\noindent Programming languages exist in a plethora of different forms and with different purposes. When deciding on programming languages we evaluate them using subjective metrics such as \textit{readability and natural semantics} \cite{gupta2004}, \textit{conciseness and simplicity} \cite{mciver1996seven, gries1974should}, and \textit{the need for specific structures in a language} \cite{milne2002difficulties}. These categorisations might be contextual to a given period, e.g., John Backus’ idea of elegant languages from 1978 significantly differs from the general idea of an elegant language today \cite{mullerRhetoricalFrameworkProgramming2020}. Yet, we do not consider adapting the language itself. We understand that some things will always be said poorly in any language \cite{steimannFatalAbstraction2018a} and accept the limitations of the language paradigm utilised \cite{vanroyHistoryOzMultiparadigm2020a,robins1988style,mcdonald2018teaching}. All of this is a natural consequence of the typically strict governance of languages and the understanding of languages as being immutable artefacts \cite{favre2005languages, whitaker1993ada, kleppeFieldSoftwareLanguage2009a}.

In this study, I challenge the idea that programming languages should be immutable artefacts and, through a literature review, provide future language engineers with a summary of existing research on language extensibility. If we stop and look, “\textit{maybe we find something important and interesting}” as said by Richard Gabriel \cite{gabrielThrowItchingPowder2014}.

\section{Background}
\label{sec:background}
\noindent In the literature, extensibility is vaguely defined. Extensibility is what “\textit{permits programming language users to define new language features,}” using definition facilities to create new notations, structures, operations, and regimes of control \cite{standish1975extensibility}. Extensibility is all that which, \textit{“allows the user to enrich the language}” \cite{hemmendinger2003extensible}. In a sense every programming language that allows subprograms is extensible, as we may define new operations through functions or procedures \cite{hemmendinger2003extensible}. Wegbreit argues that extensibility is more than addition --- it should include modification, "\textit{we need a language which can be extended, modified, and thereby tailored for use in a wide variety of application areas}" \cite{wegbreit1970studies}. Wegbreit explicates the need to be able to \textit{modify} the language, which does not seem as important in the more accepted definitions. Omitting the ability to modify the language, concepts such as expressibility are used similarly to extensibility \cite{oliveiraExtensibilityMasses2012, torgersenExpressionProblemRevisited2004, michaelsonAreThereDomain2016a}. Despite these various explanations of extensibility, the definition remains broad and close to other concepts such as expressibility. 

Instead of defining extensibility itself, Standish defines the limitations of languages' extensibility. Specifically, he argues some extensions are possible within the language (Paraphrase), some are infeasible to express (Orthophrase), and, at last, what ability to alter interpretations of expressionse exist (Metaphrase) \cite{standish1975extensibility}. Standish manages to incorporate both the broad definitions initially presented (through paraphrase and orthophrase) and Wegbreit's definition (through metaphrase).

Turning to software, extensibility is a more well-defined concept, i.e., ``\textit{software is \textit{extensible} if it can be adapted to possibly unanticipated changes in the specification}'' \cite{zenger2004programming}. In terms of software, extensibility allows for more reuse and adaptability. Thus, extensibility is an important requirement for software such as libraries and frameworks which are to be used by other programmers. It is, therefore, interesting that extensibility is much less explored in programming languages.

Further, extensibility is central to natural languages; new dialects can be created and used within the natural languages \cite{naur1975programming}. One can even formulate entirely new languages, such as the mathematical language. Similar proposals have been made for programming languages, such as Guy Steele claiming that designing for language growth, i.e., extensibility, should be essential for future languages \cite{steele1998growing}. 

While there is no clear definition of extensibility, it has been part of programming almost since its inception in the form of library catalogues for machine code \cite{wilkes1951preparation} and later macro languages for assembly languages \cite{weise1993programmable}. Extensibility has continued to influence languages through powerful module abstractions, syntactic configurability, and code isolation \cite{smaragdakisNextparadigmProgrammingLanguages2019}. Furthermore, extensibility is essential in system development, as it may be used to “future-proof” a solution \cite{wirth1988type}. Therefore, I here attempt to investigate how the research community has explored extensibility.

\subsection{Extensibility in Programming Languages}

While extensibility is present in current programming languages, it often exists as an add-on. Investigating concepts such as substitution in the Lambda-calculus shows that this is an informal meta-level construction \cite{abadi1991explicit}. Similarly, programming languages have such mechanisms that are kept as separate systems \cite{heinleinMOSTflexiPLModularStatically2012a}, e.g., C’s preprocessor \cite{lorenzenSoundTypedependentSyntactic2016a}. While this is the general pattern, some languages do incorporate and centralise extensibility, such as Lisp \cite{monnierEvolutionEmacsLisp2020}, Ruby on Rails and Emacs \cite{kingHistoryGroovyProgramming2020a},  Lua \cite{ierusalimschyEvolutionLua2007}, Lithe \cite{sandbergLitheLanguageCombining1982a}, and Langrams \cite{steimannReplacingPhraseStructure2017a}. Some language engineers also acknowledge the benefits of enabling user-defined features in languages such as Prolog \cite{colmerauer1996birth}. However, researchers still claim that extensible languages receive little attention \cite{heinleinMOSTflexiPLModularStatically2012a, steimannFatalAbstraction2018a}. Arguing this is the case as it yields less intuitive semantics \cite{felleisen1991expressive} and incomprehensible source code as a result of code expansions \cite{heinleinMOSTflexiPLModularStatically2012a, liSlimmingLanguagesReducing2015a, pombrioInferringTypeRules2018a}. A lack of extensibility may also relate to having other desired properties, such as avoiding division of the community into dialect groups \cite{stroustrupEvolvingLanguageReala, hudakHistoryHaskellBeing2007a} or avoiding bad coding styles \cite{liskov1993history}.

\subsection{Extensibility as an Isolated Concept}

While research and language engineers utilise --- and sometimes centralise --- extensibility, it is rarely directly confronted. This includes discussions of macros \cite{pombrioInferringTypeRules2018a}, reflection \cite{smith1984reflection}, functions \cite{molerHistoryMATLAB}, and modules \cite{jagannathanMetalevelBuildingBlocks1994}. While discussing these constructs, researchers primarily focus on reusability, modularity, and abstraction. These concepts are significantly more well-described than extensibility \cite{satoModuleGenerationRegret2020, spacekInheritanceSystemStructurala}. This makes them well-defined, as abstraction is defined through multiple theorems \cite{kay1993early, craryModulesAbstractionParametric2017a} and definitions of the contracts established when using abstractions \cite{stricklandContractsFirstclassModules}. Thus, extensibility remains an unexplored concept in itself.

\section{Related work}
\label{sec:related-work}
While extensibility is often not a research concern in itself, a few publications do centralise extensibility.

This includes work which attempts to improve extensibility of various languages. Daniel Zingaro applies Standish’s idea of \textit{paraphrase}, \textit{orthophrase}, and \textit{metaphrase} to a set of languages and some of their extensions \cite{zingaro2007modern}. While he reasons about extensibility, the primary focus is on improving extensibility, e.g., making macros hygienic in Scheme \cite{zingaro2007modern}. Similarly, Sebastian Erdweg and Felix Rieger present a framework for reasoning about language extensibility, but primarily they apply their system, \textit{Sugar*}, to increase extensibility \cite{erdwegFrameworkExtensibleLanguages2013}. 

Differently, Philip Wadler formulated the \textit{Expressibility problem} in the Java emailing list \cite{wadler2012}. I will not go into the details of the problem, but instead refer to other resources such as \cite{torgersenExpressionProblemRevisited2004}, \cite{oliveiraExtensibilityMasses2012}, and \cite{findler1998modular}. These papers present both the problem, but also varying solutions to it. Generally, the expressibility problem is used to measure the extensibility of a programming language. Thus, being a great tool to explore the limitations of language extensibility.

Another category of related work includes various teaching books exploring  components of programming languages. This includes the works of Robert Tennent, \textit{Principles of programming languages}, describing the different components of languages such as expressions, procedures, labels, and control structures \cite{tennent1981principles}. Furthermore, Tennent details his \textit{principle of abstraction}, defining how we may use any semantically meaningful syntactical class anywhere in our code to allow for abstract reuse. Peter Sestoft’s book, \textit{Programming Language Concepts}, also covers similar topics and discuss design decisions in widely used languages \cite{sestoft2017programming}. However, such teaching books once more focus on implementational details and discuss variations of their practical application --- not extensibility in itself.

Thus, related work focuses on improving extensibility, implementational limitations, and implementation details, while applying \textit{some} theoretical consideration. All of this suggests the need for further investigation of extensibility in itself, to allow  future language engineers to make informed choices regarding extensibility.

\section{Methodology}
\label{sec:method}

To produce an overview of approaches to extensibility, the vast amount of publications within the field of language engineering should be considered. I, therefore, conduct a literature review (from now on \textit{review}, for brevity), following the process as defined by Kitchenham et al. \cite{kitchenham2007guidelines}. This work is based on an aggregation made during the fall of 2022.

\subsection{Initial Publication Set}
I aggregated publications from venues publishing on language engineering, excluding those focusing on particular language paradigms, leaving me with 10 venues (7090 publications); \emph{HOPL}, \emph{POPL}, \emph{PLDI}, \emph{CGO}, \emph{DLS}, \emph{GPCE}, \emph{Onward}, \emph{SLE}, \emph{TOPLAS}, \emph{PACMPL}.\footnote{Includes articles from the other venues.} This excluded venues such as the ICFP and parts of OOPSLA. The consequences of the employed exclusion strategy of venues are discussed in Section \ref{sec:threat:exclusion-of-important-venues}. An overview of the initial 7090 publications can be found in Appendix \ref{app:review-traceability:unfiltered}.

\subsection{Selection Criteria and Procedure}
By collecting publications using ad hoc database queries\footnote{{e.g, ("Extensibility" OR "Extensible" OR "Syntactic sugar" OR "Difficult concepts" ) AND  ("Programming Language" OR "Language"  OR "Computer science")}} focusing on extensibility and difficulties in learning CS concepts, I identified 19 peer-reviewed publications.\footnote{\cite{irons1970experience, standish1975extensibility, wiedenbeck1999comparison, sobraltextordmasculine2021old, winograd1979beyond, milne2002difficulties, hemmendinger2003extensible, shrestha2020here, kohlbecker1986hygienic, kaczmarczyk2010identifying, piteira2013learning, tan2009learning, naur1975programming, wiedenbeck1999novice, steele1998growing, lenarcic2006antiusability,  yeomans2019transformative, zingaro2007modern, mcdonald2018teaching}} From these, I extracted 65 keywords and generated 19 additional keywords (See Appendix \ref{app:selection-criteria-keywords}). By comparing these keywords with the titles of the initial publication set, I reduced the set to 648 publications (See Appendix \ref{app:review-traceability:filter1}). Hereafter, I compared the keywords to the abstract and conclusion of the remaining publications. This further reduced the set of publications to 223 (See Appendix \ref{app:review-traceability:filter2}). Publications whose relation to the keywords are ambiguous may still be relevant when; (i) used terms are closely related to extensibility or used synonymously, as mentioned with expressibility and extensibility previously, or (ii) when an older publication describes a concept before it is formalised. Therefore, when the relevance of a publication was unclear, it remained in the set of publications.

\subsection{Quality Assessment Procedure} 
Determining the quality of the remaining 223 publications was based on whether or not a publication describes or discusses extensibility. To keep this systematic, I applied the questions found in Appendix \ref{app:quality-assessment-questions} to guide the assessment. The result was two categories of publications: Those of significant relevance (69 publications) and those of partial relevance (64 publications) (See Appendix \ref{app:review-traceability:filter3}). This approach diverges from Kitchenham, in that I do not apply their hierarchy of evidence, which focuses on bias and validity \cite{kitchenham2007guidelines}. I chose this alternative approach as bias and validity are inapplicable to some types of publications in this study, e.g., non-empirical publications based on logic and descriptions, and personal language engineer accounts, e.g., from HOPL.

\subsection*{Data Extraction}
\label{sec:analysis}
Last, I extracted and synthesised the extensibility-related contents of the 133 publications. Here I diverged from Kitchenham and instead extracted all qualitative descriptions and discussions of extensibility. I then borrowed methods from Grounded Theory (GT) to synthesise this. It is important to recognise that the continuous collection of data, which is essential to GT, is incompatible with the collection process applied here \cite{charmaz1966search}. Consequently, I do not claim to follow GT; rather, I exclusively borrow GT's data coding approach. After all, ``\textit{Grounded theory methods provide a set of different strategies for conducting rigorous qualitative research}'' \cite{charmaz1966search}. In practice, this meant an inductive approach to developing the categories I later present --- continuously improving and reevaluating categories as new information was uncovered.

\section{Results}
\label{sec:results}






During the study, four themes were discovered. The following sections elaborate on these themes. Specifically, I will detail \textit{Macros} (Section \ref{sec:results:macro}), \textit{Modules} (Section \ref{sec:results:module}), \textit{Types} (Section \ref{sec:results:type}), and \textit{Reflection} (Section \ref{sec:results:reflection}). Beyond this, two properties spanning all themes were identified, \textit{Parametricism} (Section \ref{sec:results:parametricism}) and \textit{First-class citizen behaviour} (Section \ref{sec:results:first-class}). Table \ref{tab:overview} provides an overview of the publications associated with each theme and property.

\begin{table}[!htbp]
\resizebox{\columnwidth}{!}{%
\begin{tabular}{ll}
\textbf{Macros} (Section \ref{sec:results:macro})     & \cite{weise1993programmable, dershowitz1991rewrite, williamsFlexibleNotationSyntactic1982a, holt1979model, laddadFluidQuotesMetaprogramming2020, krishnamurthiDesugaringPracticeOpportunities2015, huangMorphingStructurallyShaping2011, kohlbecker1987macro, disneySweetenYourJavaScript2014, ballantyneMacrosDomainspecificLanguages2020a, bakerMayaMultipleDispatchSyntaxa,tobinhochstadtLanguagesLibraries, monnierEvolutionEmacsLisp2020, kiselyovMacrosThatCompose2002a, erdwegFrameworkExtensibleLanguages2013, kingHistoryGroovyProgramming2020a, waddellExtendingScopeSyntactic1999a, culpepperSyntacticAbstractionComponent2005, zhangCompositionalProgramming2021, heinleinMOSTflexiPLModularStatically2012a, steele1996evolution, dos2006specifying, stroustrupEvolvingLanguageReala, suttonDesignConceptLibraries2012, stroustrupThrivingCrowdedChanging2020a} \\
\textbf{Modules} (Section \ref{sec:results:module})   & \cite{macqueenUsingDependentTypesa, holt1979model, symeEarlyHistory2020a, smaragdakisNextparadigmProgrammingLanguages2019, hickeyHistoryClojure2020, hansen1996monitors, ierusalimschyEvolutionLua2007, stricklandContractsFirstclassModules, jones1996using, blume1999hierarchical, garyWhatRecursiveModulea, liskov1993history, wirthModula2Oberon2007a, blume1999hierarchical, stroustrup1996history, dreyerTypeSystemHigherOrdera, kilpatrickBackpackRetrofittingHaskell2014, rather1996evolution, hudakHistoryHaskellBeing2007a, stroustrupEvolvingLanguageReala, monnierEvolutionEmacsLisp2020}                                                                                                                                                                                                                                                                       \\
\textbf{Types}    (Section \ref{sec:results:type})  & \cite{brookes2014essence, tofte1994principal, stroustrup1996history, stricklandContractsFirstclassModules, ierusalimschyEvolutionLua2007, watanabeProgramGenerationML2017, kay1993early, morrisonAdHocApproach, vanroyHistoryOzMultiparadigm2020a, meyer1986type, wadlerHowMakeAdhoc1989a, siekEssentialLanguageSupporta, kaminaLightweightScalableComponents2007, jansson1997polyp, wolfingerAddingGenericityPlugin, stroustrupEvolvingLanguageReala, wegnerUnificationDataProgram1983a, siekEssentialLanguageSupporta, dos2006specifying}                                                                                                                                                                                                                                                                                                                                 \\
\textbf{Reflection} (Section \ref{sec:results:reflection}) & \cite{smith1984reflection, teruelAccessControlReflection, huangExpressiveSafeStatic, politoBootstrappingInfrastructureBuild2015a, ierusalimschyEvolutionLua2007, miaoCompiletimeReflectionMetaprogramming2014, politoBootstrappingInfrastructureBuild2015a, huangMorphingStructurallyShaping2011, wirfs-brockJavaScriptFirst202020a}                                                                                                                     \\                                                              \textbf{Parametricism} (Section \ref{sec:results:parametricism}) & \cite{oderskySimplicitlyFoundationsApplications2018a,wirth1996recollections,hudakHistoryHaskellBeing2007a, servettoMetacircularLanguageActivea, tofte1994principal, symeEarlyHistory2020a, macqueenHistoryStandardML2020a, heinleinMOSTflexiPLModularStatically2012a, oderskySimplicitlyFoundationsApplications2018a, lewisImplicitParametersDynamic2000a}                                  

\\
\textbf{First-class citizen behaviour} (Section \ref{sec:results:first-class}) & 
\cite{jagannathanMetalevelBuildingBlocks1994, hansen1996monitors, wegnerUnificationDataProgram1983a, dreyerTypeSystemHigherOrdera, wirfs-brockJavaScriptFirst202020a, jones1996using, liskov1993history,  homerFirstclassDynamicTypes2019a, gelernterEnvironmentsFirstClass1987a} 

\end{tabular}%
}
\vspace{0.5em}
\caption{The publications associated with each theme and properties.}
\label{tab:overview}
\end{table}



\subsection{Macros}
\label{sec:results:macro}
\begin{motivation}
\textbf{Macros} enable language users to create new regimes of control of almost arbitrary syntactically form, hence increasing extensibility.
\end{motivation}

\noindent Macros were introduced along assembly languages \cite{weise1993programmable}.  Macros are typically categorised as \textit{term rewriting}; some transformation rules are applied until the system reaches a terminal \cite{dershowitz1991rewrite}. This is also called \textit{string productions} \cite{williamsFlexibleNotationSyntactic1982a}. A benefit of macro systems, as a consequence of the application of rules, is that it guarantees to reach something irreducible, i.e., reducing to constants \cite{holt1979model}. However, macros have severe consequences.  This includes the introduction of significant code bloat as a result of string expansions \cite{laddadFluidQuotesMetaprogramming2020}, also making the produced code difficult to debug \cite{krishnamurthiDesugaringPracticeOpportunities2015}. It typically brings consequences of not being analysable \cite{huangMorphingStructurallyShaping2011} and difficult to implement for complex languages \cite{weise1993programmable}. Furthermore, macros establish some implementational problems, especially that of hygienic macros is well-studied, i.e., ensuring that different macros do not interfere with each other \cite{kohlbecker1987macro, disneySweetenYourJavaScript2014, ballantyneMacrosDomainspecificLanguages2020a, bakerMayaMultipleDispatchSyntaxa,tobinhochstadtLanguagesLibraries, monnierEvolutionEmacsLisp2020, kiselyovMacrosThatCompose2002a}.

The investigated publications define three types of macros: (i) \textit{lexical}, (ii) \textit{syntactical}, and (iii) \textit{semantical}. Lexical macros simply apply string-based transformations. A benefit and consequence of lexical macros are their syntax-agnostic approach; they do not guarantee the preservation of any language syntax  \cite{erdwegFrameworkExtensibleLanguages2013}. Syntactical macros operate on the abstract syntax trees (ASTs) of a language to guarantee syntactical safety \cite{weise1993programmable}. Such macros can be more cumbersome to write as they use a format foreign to the programmer. Consequently, languages such as Groovy utilise automatic expansions from code to ASTs to simplify writing macros \cite{kingHistoryGroovyProgramming2020a}. Common for both lexical and syntactical macros are that they provide no semantic analysis, such as type checking --- which is the case for semantic macros \cite{weise1993programmable}.

Macros are highlighted as significantly powerful in combination with modules \cite{waddellExtendingScopeSyntactic1999a} because this yields greater expressibility by allowing users to define macros as modules \cite{waddellExtendingScopeSyntactic1999a,culpepperSyntacticAbstractionComponent2005}. This is, among other things, an interesting feature of the Racket system \cite{tobinhochstadtLanguagesLibraries}. Beyond these comments, I highlight the following contents of the studied literature:

\begin{itemize}
\item Racket, which features a modular syntactical system \cite{zhangCompositionalProgramming2021}.
\item CPP, the C language’s simple lexical macro engine \cite{zhangCompositionalProgramming2021}.
\item Dylan’s rewrite macro system that utilises a procedural macro engine \cite{heinleinMOSTflexiPLModularStatically2012a}.
\item Maya where grammar productions act like compile-time generic functions \cite{heinleinMOSTflexiPLModularStatically2012a}.
\item FlexiPL allowing the addition of completely new syntax without sacrificing static type safety \cite{heinleinMOSTflexiPLModularStatically2012a}.
\item LISPs (and other LISP languages’) syntactical macro system \cite{steele1996evolution,monnierEvolutionEmacsLisp2020}, with proposals for lexical scope extensions  \cite{culpepperSyntacticAbstractionComponent2005}.
\item C++ templates and concepts, which albeit not being macros, are macro-like \cite{dos2006specifying, stroustrupEvolvingLanguageReala, suttonDesignConceptLibraries2012, stroustrupThrivingCrowdedChanging2020a}.
\end{itemize}


\subsection{Modules}
\label{sec:results:module}
\begin{motivation}
\textbf{Modules} are essential to structure code, allowing users to group code such that they can provide domain-specific languages or frameworks as a package. Thus, modules are a significant tool for structuring and presenting extensibility.
\end{motivation}

\noindent In programming languages, users package, cluster, and modularise their code. They partition it to create manageable components, which are connected to create programs \cite{macqueenUsingDependentTypesa, holt1979model}. I will collectively call various programming concepts enabling such features 'modules'. Today, modules have become essential to programming \cite{symeEarlyHistory2020a, smaragdakisNextparadigmProgrammingLanguages2019}. A lack of existing modules is often associated with a lack of power \cite{hickeyHistoryClojure2020}. However, modules do introduce their own issues, some researchers argue that modularity enforces the need for more documentation \cite{hansen1996monitors}. Module concepts should also be considered early in language design, as implementing this or changing it post-mortem can have severe consequences \cite{ierusalimschyEvolutionLua2007}.

Furthermore, it becomes important to design by contracts as the contents of a module are not necessarily known \cite{stricklandContractsFirstclassModules}. The contract is often upheld using type systems, and as such, multiple types have been suggested for modules; (i) \textit{existential}, (ii) \textit{dependent}, and (iii) \textit{manifest} types. Existential types allow first-class modules but hide almost all implementational details. Opposite of this, dependent types are completely transparent. Lastly, manifest types provide a compromise between the two other types \cite{jones1996using}. While an oversimplification, one can think of these levels as; (i) relying on a module in a dynamic language, (ii) using a module directly in a statically typed language, and (iii) implementing a module against interfaces in a statically typed language.

When developing modules, one may also rely on other modules. This hierarchical approach is essential to developing with modules \cite{blume1999hierarchical, garyWhatRecursiveModulea}. Consequently, it is important to reason about hierarchies and dependencies. One concern is whether or not nested modules should be kept separate \cite{liskov1993history, wirthModula2Oberon2007a} or compiled separately \cite{blume1999hierarchical, stroustrup1996history}.\footnote{Separate compilation is a well-investigated concept, with many proposed solutions, such as \cite{dreyerTypeSystemHigherOrdera}.} Furthermore, what should happen when multiple modules rely on the same module \cite{kilpatrickBackpackRetrofittingHaskell2014}? How are circular dependencies handled? These are all questions language engineers should consider. Beyond these thoughts, I highlight the following contents of the investigated literature:

\begin{itemize}
	\item The Forth language as a whole, where everything is a module \cite{rather1996evolution}.
	\item The Haskell module construct which simply exports algebraic types \cite{hudakHistoryHaskellBeing2007a}.
	\item The C++ module construct which uses a single namespace mechanism \cite{stroustrupEvolvingLanguageReala}.
	\item The Emacs language where code loosely follows a global naming convention \cite{monnierEvolutionEmacsLisp2020}.
\end{itemize}

\subsection{Types}
\label{sec:results:type}
\begin{motivation}
\textbf{Type systems} allow users to make guarantees in languages. Typically, type constructs restrict extensibility. However, type systems also increases the predictability of a language. Consequently, types, combined with properties such as polymorphism and generic programming, enable the predictability of user-developed programs (or language extensions) and thereby provide safety concerning extensibility.
\end{motivation}

\noindent Generally, types are more than error checking  \cite{brookes2014essence}. Types allow us to assert equality \cite{tofte1994principal, stroustrup1996history}. Consequently, types allow us to design by contracts \cite{stricklandContractsFirstclassModules}. Specifically, types are the constructs used to define what the results of expressions and statements should be either statically or dynamically typed \cite{ierusalimschyEvolutionLua2007}.  Type constructs and systems are not necessarily wanted. Types are perceived to limit expressiveness \cite{watanabeProgramGenerationML2017}. Properties of type systems, such as inheritance, are also by some researchers argued to be limiting \cite{kay1993early}.

While I here provide a brief introduction to types, there exists significantly more literature on the topic. However, as the goal of this study is not to map publications on the topic of types alone, I will focus on that which appears most relevant to extensibility. Specifically, this is polymorphism and generic programming.  

Originally, the term polymorphism was used for parametric functions \cite{kay1993early}. However, today, polymorphism is that which allows one to operate over multiple types of data, allowing more general reuse of a piece of code \cite{morrisonAdHocApproach}. Generally, polymorphism is what adds flexibility to a type construct \cite{vanroyHistoryOzMultiparadigm2020a}. It even allows for interesting behaviour such as recursive polymorphism \cite{meyer1986type}. 

Polymorphism is divided into two categories: (i) \textit{ad hoc} and (ii) \textit{universal}. While an operational definition exists for ad hoc polymorphism, i.e., that the type of the code evaluation depends on its arguments, there exists no single systematic approach \cite{morrisonAdHocApproach}. Universal polymorphism contains both parametric polymorphism and subtype polymorphism. Parametric polymorphism allows the use of any type as a parameter. Alternatively, subtype polymorphism allows the substitution of a type with any type contractually having the same capabilities.\footnote{This may be determined by inheritance or structural matching depending on the language implementation of subtype polymorphism.} The key difference between ad hoc and universal polymorphism is that universal polymorphism executes the same code for all (possible) types.

Some languages allow for both ad hoc and universal polymorphism \cite{morrisonAdHocApproach}. Multiple researchers suggest that type classes, as found in Haskell, may ease the use of polymorphism, e.g., \cite{wadlerHowMakeAdhoc1989a} and  \cite{siekEssentialLanguageSupporta}. The literature also details these related terms:

\begin{itemize}
\item \textbf{family polymorphism}. This allows users to create mutually recursive classes, which are named a family. These can safely be extended without changing the code. Family polymorphism, its problems, and potential solutions are further described in \cite{kaminaLightweightScalableComponents2007}.

\item \textbf{Polyptic functions}. These are more general than a polymorphic function. This specifically relates to the continuous repetitive implementations of functions for different data types. These are described in \cite{jansson1997polyp}.
\end{itemize}

\noindent Polymorphism allows for generic programming; working with generic types, or at least generic until specified later. The Ada language pioneered this. Genericity exists in many forms. In some languages, genericity is a concept supported in the virtual machine, e.g., C\#. In contrast, other languages eliminate the generic types, e.g., Java, where genericity is solved by \textit{object} references at runtime \cite{wolfingerAddingGenericityPlugin}. An interesting feature of polymorphism and generic programming is that it is paradigm independent  \cite{stroustrupEvolvingLanguageReala}. The specific implementation of generic programming exists in many forms; CLU provides one step for declaration of generic parts, and one for instantiation  \cite{wegnerUnificationDataProgram1983a} and further relies on structural matching, i.e., requiring the contents of types to be the same. Structural matching is also what ML, C++, and  Ocaml utilise \cite{siekEssentialLanguageSupporta}. 

Beyond what has already been mentioned, the literature  highlights the type construct of C++.  More specifically, C++ concepts and templates for generic programming \cite{dos2006specifying}.

\subsection{Reflection}
\label{sec:results:reflection}
\begin{motivation}
\textbf{Reflection} enables users to modify other programming constructs programmatically. As such, this increases modifiability and flexibility of concepts otherwise inaccessible to modification, which results in an increase in extensibility.
\end{motivation}

\noindent Reflection is a powerful feature that allows programs to inspect and (sometimes) modify themselves \cite{smith1984reflection,teruelAccessControlReflection}. Some utilise reflection to clone and extend classes \cite{huangExpressiveSafeStatic}. Research also argues that reflective languages should not only be able to reason about themselves but also their surroundings \cite{smith1984reflection}. Reflection can also introduce a clear interface for the virtual machine \cite{politoBootstrappingInfrastructureBuild2015a}. Often, reflection is thought of as requiring meta-objects, or constructs. However, when reflecting upon user-developed components, the meta facilities already exist, i.e., the programming constructs used to build this composite  \cite{ierusalimschyEvolutionLua2007}. While reflection is powerful, runtime-based reflection often comes with a performance cost and breaks encapsulation \cite{miaoCompiletimeReflectionMetaprogramming2014}. Furthermore, it may yield metastability problems, i.e., cause meta-call recursion rendering the system unusable \cite{politoBootstrappingInfrastructureBuild2015a}. Beyond these ideas, I highlight the following contents of the investigated literature:

\begin{itemize}
	\item The LISP reflective construct. However, this is highlighted as an argument that there needs to be a meta-representation of the system, as there is more to reasoning than just referencing --- one needs a theory \cite{smith1984reflection}.
	\item MorphJ which adds reflective iteration blocks to Java \cite{huangMorphingStructurallyShaping2011}.
	\item Javascript’s history provides some thoughts on different directions for a reflective system, as given in \cite{wirfs-brockJavaScriptFirst202020a}.
\end{itemize}

\subsection{Cross-theme Properties}
\label{sec:results:properties}

Beyond the highlighted themes, literature also presents properties which can be associated with the various themes, which in turn increases extensibility of a language. I here highlight the two most referred to by literature.

\subsubsection{Parametricism}
\label{sec:results:parametricism}

\begin{motivation}
\textbf{Parametrisicm} increases the customizability of constructs, enabling further reuse and abstraction. Consequently, parametrizing programming constructs increases flexibility of such constructs, and thereby extensibility of the language. 
\end{motivation}

\noindent Parameters allow postponing the specifics of utilised variables -- eliminating the need to handle dependencies  \cite{oderskySimplicitlyFoundationsApplications2018a}. However, parametricism is not only applicable to functions, it can also be applied to other constructs, e.g., modules and types.  The parametricism of a construct may vary, e.g., pass by value or by reference. Both types have their own benefits \cite{wirth1996recollections}. In some languages either is used exclusively, but some languages also differ depending on the type of parameter. The Java language, as an example, uses pass-by-value with primitive types, and pass-by-reference with object types. Other languages allow the user to choose, e.g., C\# does so by the \textit{ref} keyword.\footnote{C\# allows even greater variation, such as \textit{in} and \textit{out} parameters. These can be found here: \url{https://docs.microsoft.com/en-us/dotnet/csharp/language-reference/keywords/}.} Another variation is the use of named parameters \cite{homerAPIsLanguagesGeneralising2015a}. Named parameters allow one to provide parameters using names, increasing the visibility of a parameter’s purpose.  Lastly, parameters may be optional to provide flexibility in use, e.g., as in JavaScript\footnote{\url{https://developer.mozilla.org/en-US/docs/Web/JavaScript/Reference/Functions/Default_parameters}} or C\#.\footnote{\url{https://docs.microsoft.com/en-us/dotnet/csharp/programming-guide/classes-and-structs/named-and-optional-arguments}}
However, using parameterisation increases the complexity of a construct  \cite{hudakHistoryHaskellBeing2007a}. As I will detail below, the investigated literature discusses parametricism in practice for both \texttt{modules} and \texttt{types}.
\\\\
\noindent\textbf{Modules}
\\
\noindent One idea is \textit{Active modules}. These are parametric since one can supply parameters and, based on these, receive a customised variation of the module \cite{servettoMetacircularLanguageActivea, tofte1994principal}. An example of parametric modules, is the Standard ML and OCaml \texttt{functor} construct which can be utilised to make modules parametric \cite{symeEarlyHistory2020a, macqueenHistoryStandardML2020a}. The OCaml documentation provides a tutorial for this.\footnote{\url{https://caml.inria.fr/pub/docs/oreilly-book/html/book-ora132.html}}
\\\\
\noindent\textbf{Types}
\\
\noindent Types may also be parametric. Specifically, polymorphism may function as parametricism for types as is the case with parametric polymorphism. However, other types of polymorphism further increase the flexibility of parametricism on types. For example, one can qualify the generic parameters by placing certain bounds on them, e.g., forcing them to extend a given interface (subtype polymorphism). examples hereof include the use of concepts to bind templates in  C++\footnote{\url{https://en.cppreference.com/w/cpp/language/constraints}}, and type classes in Haskell.\footnote{\url{https://www.haskell.org/tutorial/classes.html}} 

While not specifically related to parametric types, parametric properties and type constructs together allow for features such as deduced or implicit parameters \cite{heinleinMOSTflexiPLModularStatically2012a, oderskySimplicitlyFoundationsApplications2018a}. Specifically, this allows the system to deduce which parameter should be used. Implicit parameters are a way of dealing with too many parameters, thus reducing the complexity of user-developed constructs \cite{lewisImplicitParametersDynamic2000a}. 

\subsubsection{First-class citizen behaviour}
\label{sec:results:first-class}
\begin{motivation}
The property of being \textbf{first-class} enables the user to use a construct as any other value, allowing modification of the programming construct dynamically. This implies complete freedom of modification within the current language, removing most impediments of it being solidified in the language. Consequently, this increases extensibility similarly to parametricism but yields greater flexibility as modifications can be more than changing parameters.
\end{motivation}
\noindent First-class citizens is the term used to describe values of a system which we can operate on, e.g., modify, copy, pass as values \cite{scott2000programming}. Constructs being first-class yields significantly more flexibility as they can be treated like all other values in a program. Once more, the construct property of being \textit{first-class} is discussed both in regards to \texttt{modules} and \texttt{types}. 
\\\\
\noindent\textbf{Modules}
\\
\noindent Researchers argue that modules should be data structures which can be operated upon like any other value \cite{jagannathanMetalevelBuildingBlocks1994}. They question why modules cannot be used in such a way --- it appears as an arbitrary choice \cite{hansen1996monitors, wegnerUnificationDataProgram1983a}. Making modules first-class would increase their modifiability and thereby extensibility \cite{dreyerTypeSystemHigherOrdera}. However, keeping modules second-class is also used as a way of keeping programs safe \cite{wegnerUnificationDataProgram1983a}. As such, first-class modules allow more flexibility in creating new programming constructs, whereas second-class modules primarily act as ways of structuring code \cite{wirfs-brockJavaScriptFirst202020a}. An example of extending a module construct to be first-class (in LISP), can be found in \cite{gelernterEnvironmentsFirstClass1987a}. Furthermore, \cite{jones1996using} illustrates how polymorphism can be used to create first-class modules.
\\\\
\noindent\textbf{Types}
\\
\noindent Types may also be first-class. Specifically, \cite{homerFirstclassDynamicTypes2019a} provides an example of this. Another way of doing so is using type generators. Type generators are not types themselves but allow the generation of types. In CLU this is how arrays are implemented, i.e., arrays are not types, but providing the array (type-generator) construct with a type, e.g., \textit{int},  “instantiates” the type \cite{liskov1993history}. Type generators are also more broadly used in languages such as F\#\footnote{\url{https://docs.microsoft.com/en-us/dotnet/fsharp/tutorials/type-providers/}} or Scala (via its macro construct).\footnote{\url{https://docs.scala-lang.org/overviews/macros/typeproviders.html}}

\subsection{Less Explored Concepts}
\label{sec:results:less-explored}
Some language features --- not necessarily constructs we apply directly --- make extensibility more complex by blurring the line between language and language platform. Such features should be of interest to language engineers as these also enable unique extensibility mechanisms. This section will introduce the identified features, which are generally less explored within the identified research.
\\\\
\noindent \textit{Bootstrapping}. Bootstrapping is the initialisation of a language. Specifically, it is the process that produces the necessary programs to execute a language. Circular bootstrapping is the process wherein a language bootstraps itself  \cite{politoBootstrappingInfrastructureBuild2015a}. Circular bootstrapping eliminates the idea that the language cannot be modified, as the language is now written in terms of itself. However, as the language executables are still (a) separate program(s) the language is not necessarily open for modification.
\\\\
\noindent \textit{Compiler hooks}. Beyond circular bootstrapping, some languages allow users to inject hooks into the compiler, written in the language itself, significantly enhancing extensibility. This is the case in the Groovy language \cite{kingHistoryGroovyProgramming2020a} and the Java language, as exemplified by the Lombok project.\footnote{\url{https://projectlombok.org}} The Forth language also stores user-defined words in the same way as compile-time words, thereby enabling bootstrapping features at the very core of the language \cite{steimannReplacingPhraseStructure2017a}. Such features blur the line between the language and the language executables, as any program can alter the language. Similarly, some languages allow the user to escape the current language and make calls to other languages, e.g., Lua with \textit{lua\_register} and Python with \textit{bindings}. This challenges the fundamental idea of where the language begins and where the language executables end. Such features are therefore interesting subjects for future work, as they inform a unique way of increasing extensibility with only limited coverage in current publications.
\\\\
\noindent\textit{Virtual machines}. Virtual machines provide rich features and portability. New languages are often built on top of existing virtual machines, e.g., Scala, Groovy, and Clojure, which are built on the Java virtual machine.\footnote{Clojure is also available on top of other languages such as Javascript.} Generally, virtual machines inspire language experimentation \cite{tobinhochstadtLanguagesLibraries}. While such virtual machines are themselves languages --- and thus not particularly unique in regards to extensibility --- it is interesting how they are used for language development. 
Investigating the extensibility benefits of VMs and their use is also not covered by the investigated literature and may therefore be relevant to future research.

\section{Discussion}
\label{sec:discussion}


The entirety of this study has focused on providing research on extensibility, highlighting the opportunities and benefits associated with customized language use. However, as with software development, language design is a holistic process wherein one should retain multiple qualities. As stated by Glaze and Van Horn, “\textit{The strength of a dynamic language is also its weakness: run-time flexibility comes at the cost of compile-time predictability}” \cite{glazeAbstractingAbstractControl2014b}. One should, therefore, carefully consider the degree of extensibility in any language. Beyond compile-time predictability, extensibility poses multiple challenges.
 
Extensibility may reduce performance, e.g., the abstraction and extensibility opportunities of languages like Java and C\# makes them less performant than the C language \cite{altmanKeynoteLanguageOptimizer2011a}. Researchers also argue that constructs such as those for concurrency should be built into the language, to guarantee the correctness of the code \cite{boehmThreadsCannotBea}. 
Furthermore, shared user-developed constructs, e.g., community-developed modules, are typically not given complete specifications, leaving the end user to explore the modules themselves \cite{parkinsonSeparationLogicAbstractiona}. The ability to include any user-developed extensions also suggests security vulnerabilities as one can use construct developed by untrustworthy developers \cite{fournetFullyAbstractCompilationa}. 
Additionally, such extensions are typically introduced to solve problem-specific obstacles, rendering them unusable in other contexts and potentially bloating the language \cite{steimannFatalAbstraction2018a}. The very point of extension, and even abstractions, is that they should be self-contained so that one can use them without knowing the implementational details. However, in most languages, such abstractions leak, e.g., one needs to understand implementational details of integers in Java to avoid overflowing \cite{steimannFatalAbstraction2018a}.

These aforementioned caviats of extensibility might be why its influence is lesser than other qualities; allowing only specific types of language extensions implies stronger control and security of the system and thus better reliability. Such strict requirements, and uniform ways of programming, are necessary for particular problems and thereby also languages, as was the case with ALGOL60 and ADA \cite{whitaker1993ada,naurPAPEREUROPEANSIDEa}.

However, the benefits of extensibility cannot be neglected. Extensibility and the capability to create the necessary composite constructs is the only way to work around lacking parts of a language \cite{brightOriginsProgrammingLanguage2020}; the availability of community-driven extensions is what boosts developer productivity \cite{lamPuttingSemanticsSemantic2020a} and the ability to recreate certain extensions, e.g., using design patterns or macros, is what allows users to circumvent flaws  \cite{chambersDebateLanguageTool2000a}.

Because of these benefits, extensibility should be an integrated part of any holistic language design. However, when doing so, language engineers should consider the challenges highlighted.

\section{Limitations}
\label{sec:ttv}

In the methodology section, I presented several threats to the study. I will here elaborate upon these.

\subsection{Exclusion of important venues}
\label{sec:threat:exclusion-of-important-venues}
The identification of work such as  \cite{oliveiraExtensibilityMasses2012} from ECOOP, which was related to solving the extensibility problem, suggests that the scope of the chosen venues could be too narrow. The initial set of publications was 7090. However, including paradigm-specific venues and other venues on programming, e.g., OOPSLA and ICFP, give a set much closer to 14000 publications. As such, one could extend this set to cover venues more broadly in future research. However, the initial publication set was already large and exhausted the included venues. It was within this set I also identified the expressibility problem initially, which in turn enabled me to identify related work such as  \cite{oliveiraExtensibilityMasses2012}. \textbf{However, I do not claim this study to be exhaustive in any shape or form. }

\subsection{Researcher bias}
\label{sec:threat:researcher-bias}
As I solemnly conducted the review of articles, it poses a threat of bias. However, multiple steps have been taken to counter this. Kitchenham et al. suggest multiple reviewers \cite{kitchenham2007guidelines}. While this was not the case for this study, I heavily relied on my colleagues to define the exclusion strategy and quality assessment protocol. Furthermore, I leave traces of the different stages in Appendix \ref{app:review-traceability}. 
While all of these are attempts at creating a reproducible non-biased study, it still leaves the possibility of different biases, e.g., favouring HOPL papers, focusing on non-technical descriptions.

\section{Conclusion}
\label{sec:conclusion}
In conclusion, this study delves into the nuanced realm of programming language extensibility, shedding light on aspects often sidelined in language design. Through a literature review, key extensibility themes - Macros, Modules, Types, and Reflection - are identified, showcasing diverse strategies for allowing extensibility. The examination extends to cross-theme properties like Parametricism and First-class citizen behaviour, adding complexity by emphasizing the significance of customizability and flexibility in programming language constructs.

This research underscores the essential role of extensibility in overcoming language limitations and fostering community-driven development. 
While this study underlines central themes of extensibility, it also suggests further exploration of other fundamental concepts such as bootstrapping, compiler hooks, and virtual machines, recognizing their potential impact on language extensibility.

While advocating for an increased focus on language extensibility, it's important to acknowledge challenges such as performance considerations, security vulnerabilities, and the potential for language bloat. The study also acknowledges limitations, including biases in venue selection. Nonetheless, by consolidating years of research, this work provides a foundational resource for future language designers venturing into the realm of extensibility.

\bibliographystyle{unsrt}
\bibliography{bibliography}

\begin{thebibliography}{100}

\bibitem{gab2012}
Richard~P. Gabriel.
\newblock The structure of a programming language revolution.
\newblock In {\em Proceedings of the {{ACM}} International Symposium on {{New}}
  Ideas, New Paradigms, and Reflections on Programming and Software -
  {{Onward}}! '12}, page 195, {Tucson, Arizona, USA}, 2012. {ACM Press}.

\bibitem{gupta2004}
Diwaker Gupta.
\newblock What is a good first programming language?
\newblock {\em Crossroads}, 10(4):7--7, 2004.

\bibitem{mciver1996seven}
Linda McIver and Damian Conway.
\newblock Seven deadly sins of introductory programming language design.
\newblock In {\em Proceedings 1996 International Conference Software
  Engineering: Education and Practice}, pages 309--316. IEEE, 1996.

\bibitem{gries1974should}
David Gries.
\newblock What should we teach in an introductory programming course?
\newblock In {\em Proceedings of the fourth SIGCSE technical symposium on
  Computer science education}, pages 81--89, 1974.

\bibitem{milne2002difficulties}
Iain Milne and Glenn Rowe.
\newblock Difficulties in learning and teaching programming—views of students
  and tutors.
\newblock {\em Education and Information technologies}, 7(1):55--66, 2002.

\bibitem{mullerRhetoricalFrameworkProgramming2020}
Stefan~K. Muller and Hannah Ringler.
\newblock A rhetorical framework for programming language evaluation.
\newblock In {\em Proceedings of the 2020 {{ACM SIGPLAN International
  Symposium}} on {{New Ideas}}, {{New Paradigms}}, and {{Reflections}} on
  {{Programming}} and {{Software}}}, pages 187--194, {Virtual USA}, November
  2020. {ACM}.

\bibitem{steimannFatalAbstraction2018a}
Friedrich Steimann.
\newblock Fatal abstraction.
\newblock In {\em Proceedings of the 2018 {{ACM SIGPLAN International
  Symposium}} on {{New Ideas}}, {{New Paradigms}}, and {{Reflections}} on
  {{Programming}} and {{Software}}}, pages 125--130, {Boston MA USA}, October
  2018. {ACM}.

\bibitem{vanroyHistoryOzMultiparadigm2020a}
Peter Van~Roy, Seif Haridi, Christian Schulte, and Gert Smolka.
\newblock A history of the {{Oz}} multiparadigm language.
\newblock {\em Proceedings of the ACM on Programming Languages}, 4(HOPL):1--56,
  June 2020.

\bibitem{robins1988style}
Gabriel Robins.
\newblock {\em On Style, Expressibility, and Efficiency in Functional
  Programming Languages}.
\newblock UCLA, Computer Science Department, 1988.

\bibitem{mcdonald2018teaching}
Carlton McDonald.
\newblock Why is teaching programming difficult?
\newblock In {\em Higher Education Computer Science}, pages 75--93. Springer,
  2018.

\bibitem{favre2005languages}
J-M Favre.
\newblock Languages evolve too! changing the software time scale.
\newblock In {\em Eighth International Workshop on Principles of Software
  Evolution (IWPSE'05)}, pages 33--42. IEEE, 2005.

\bibitem{whitaker1993ada}
William~A Whitaker.
\newblock Ada—the project: The dod high order language working group.
\newblock {\em ACM SIGPLAN Notices}, 28(3):299--331, 1993.

\bibitem{kleppeFieldSoftwareLanguage2009a}
Anneke Kleppe.
\newblock The {{Field}} of {{Software Language Engineering}}.
\newblock In Dragan Ga{\v s}evi{\'c}, Ralf L{\"a}mmel, and Eric Van~Wyk,
  editors, {\em Software {{Language Engineering}}}, volume 5452, pages 1--7.
  {Springer Berlin Heidelberg}, {Berlin, Heidelberg}, 2009.

\bibitem{gabrielThrowItchingPowder2014}
Richard~P. Gabriel.
\newblock I {{Throw Itching Powder}} at {{Tulips}}.
\newblock In {\em Proceedings of the 2014 {{ACM International Symposium}} on
  {{New Ideas}}, {{New Paradigms}}, and {{Reflections}} on {{Programming}} \&
  {{Software}}}, pages 301--319, {Portland Oregon USA}, October 2014. {ACM}.

\bibitem{standish1975extensibility}
Thomas~A Standish.
\newblock Extensibility in programming language design.
\newblock In {\em Proceedings of the May 19-22, 1975, national computer
  conference and exposition}, pages 287--290, 1975.

\bibitem{hemmendinger2003extensible}
David Hemmendinger.
\newblock Extensible language.
\newblock In {\em Encyclopedia of Computer Science}, pages 691--692. 2003.

\bibitem{wegbreit1970studies}
Ben Wegbreit.
\newblock {\em Studies in extensible programming languages}.
\newblock PhD thesis, Harvard University, 1970.

\bibitem{oliveiraExtensibilityMasses2012}
Bruno C. d.~S. Oliveira and William~R. Cook.
\newblock Extensibility for the {{Masses}}.
\newblock In David Hutchison, Takeo Kanade, Josef Kittler, Jon~M. Kleinberg,
  Friedemann Mattern, John~C. Mitchell, Moni Naor, Oscar Nierstrasz,
  C.~Pandu~Rangan, Bernhard Steffen, Madhu Sudan, Demetri Terzopoulos, Doug
  Tygar, Moshe~Y. Vardi, Gerhard Weikum, and James Noble, editors, {\em
  {{ECOOP}} 2012 \textendash{} {{Object}}-{{Oriented Programming}}}, volume
  7313, pages 2--27. {Springer Berlin Heidelberg}, {Berlin, Heidelberg}, 2012.

\bibitem{torgersenExpressionProblemRevisited2004}
Mads Torgersen.
\newblock The {{Expression Problem Revisited}}.
\newblock In Takeo Kanade, Josef Kittler, Jon~M. Kleinberg, Friedemann Mattern,
  John~C. Mitchell, Moni Naor, Oscar Nierstrasz, C.~Pandu~Rangan, Bernhard
  Steffen, Madhu Sudan, Demetri Terzopoulos, Dough Tygar, Moshe~Y. Vardi,
  Gerhard Weikum, and Martin Odersky, editors, {\em {{ECOOP}} 2004
  \textendash{} {{Object}}-{{Oriented Programming}}}, volume 3086, pages
  123--146. {Springer Berlin Heidelberg}, {Berlin, Heidelberg}, 2004.

\bibitem{michaelsonAreThereDomain2016a}
Greg Michaelson.
\newblock Are there {{Domain Specific Languages}}?
\newblock In {\em Proceedings of the 1st {{International Workshop}} on {{Real
  World Domain Specific Languages}}}, pages 1--3, {Barcelona Spain}, March
  2016. {ACM}.

\bibitem{zenger2004programming}
Matthias Zenger.
\newblock Programming language abstractions for extensible software components.
\newblock Technical report, EPFL, 2004.

\bibitem{naur1975programming}
Peter Naur.
\newblock Programming languages, natural languages, and mathematics.
\newblock {\em Communications of the ACM}, 18(12):676--683, 1975.

\bibitem{steele1998growing}
Guy~L Steele~Jr.
\newblock Growing a language.
\newblock In {\em Object-Oriented Programming, Systems, Languages, and
  Applications (OOPSLA)}. Citeseer, 1998.

\bibitem{wilkes1951preparation}
Maurice~Vincent Wilkes, David~J Wheeler, and Stanley Gill.
\newblock {\em The Preparation of Programs for an Electronic Digital Computer:
  With special reference to the EDSAC and the Use of a Library of Subroutines}.
\newblock Addison-Wesley Press, 1951.

\bibitem{weise1993programmable}
Daniel Weise and Roger Crew.
\newblock Programmable syntax macros.
\newblock In {\em Proceedings of the ACM SIGPLAN 1993 conference on Programming
  language design and implementation}, pages 156--165, 1993.

\bibitem{smaragdakisNextparadigmProgrammingLanguages2019}
Yannis Smaragdakis.
\newblock Next-paradigm programming languages: What will they look like and
  what changes will they bring?
\newblock In {\em Proceedings of the 2019 {{ACM SIGPLAN International
  Symposium}} on {{New Ideas}}, {{New Paradigms}}, and {{Reflections}} on
  {{Programming}} and {{Software}}}, pages 187--197, {Athens Greece}, October
  2019. {ACM}.

\bibitem{wirth1988type}
Niklaus Wirth.
\newblock Type extensions.
\newblock {\em ACM Transactions on Programming Languages and Systems (TOPLAS)},
  10(2):204--214, 1988.

\bibitem{abadi1991explicit}
Martin Abadi, Luca Cardelli, P-L Curien, and J-J L{\'e}vy.
\newblock Explicit substitutions.
\newblock {\em Journal of functional programming}, 1(4):375--416, 1991.

\bibitem{heinleinMOSTflexiPLModularStatically2012a}
Christian Heinlein.
\newblock {{MOST}}-{{flexiPL}}: Modular, statically typed, flexibly extensible
  programming language.
\newblock In {\em Proceedings of the {{ACM}} International Symposium on {{New}}
  Ideas, New Paradigms, and Reflections on Programming and Software -
  {{Onward}}! '12}, page 159, {Tucson, Arizona, USA}, 2012. {ACM Press}.

\bibitem{lorenzenSoundTypedependentSyntactic2016a}
Florian Lorenzen and Sebastian Erdweg.
\newblock Sound type-dependent syntactic language extension.
\newblock In {\em Proceedings of the 43rd {{Annual ACM SIGPLAN}}-{{SIGACT
  Symposium}} on {{Principles}} of {{Programming Languages}}}, pages 204--216,
  {St. Petersburg FL USA}, January 2016. {ACM}.

\bibitem{monnierEvolutionEmacsLisp2020}
Stefan Monnier and Michael Sperber.
\newblock Evolution of {{Emacs Lisp}}.
\newblock {\em Proceedings of the ACM on Programming Languages}, 4(HOPL):1--55,
  June 2020.

\bibitem{kingHistoryGroovyProgramming2020a}
Paul King.
\newblock A history of the {{Groovy}} programming language.
\newblock {\em Proceedings of the ACM on Programming Languages}, 4(HOPL):1--53,
  June 2020.

\bibitem{ierusalimschyEvolutionLua2007}
Roberto Ierusalimschy, Luiz~Henrique {de Figueiredo}, and Waldemar Celes.
\newblock The evolution of {{Lua}}.
\newblock In {\em Proceedings of the Third {{ACM SIGPLAN}} Conference on
  {{History}} of Programming Languages}, {San Diego California}, June 2007.
  {ACM}.

\bibitem{sandbergLitheLanguageCombining1982a}
David Sandberg.
\newblock Lithe: A language combining a flexible syntax and classes.
\newblock In {\em Proceedings of the 9th {{ACM SIGPLAN}}-{{SIGACT}} Symposium
  on {{Principles}} of Programming Languages - {{POPL}} '82}, pages 142--145,
  {Albuquerque, Mexico}, 1982. {ACM Press}.

\bibitem{steimannReplacingPhraseStructure2017a}
Friedrich Steimann.
\newblock Replacing phrase structure grammar with dependency grammar in the
  design and implementation of programming languages.
\newblock In {\em Proceedings of the 2017 {{ACM SIGPLAN International
  Symposium}} on {{New Ideas}}, {{New Paradigms}}, and {{Reflections}} on
  {{Programming}} and {{Software}}}, pages 30--43, {Vancouver BC Canada},
  October 2017. {ACM}.

\bibitem{colmerauer1996birth}
Alain Colmerauer and Philippe Roussel.
\newblock The birth of prolog.
\newblock In {\em History of programming languages---II}, pages 331--367. 1996.

\bibitem{felleisen1991expressive}
Matthias Felleisen.
\newblock On the expressive power of programming languages.
\newblock {\em Science of computer programming}, 17(1-3):35--75, 1991.

\bibitem{liSlimmingLanguagesReducing2015a}
Junsong Li, Justin Pombrio, Joe~Gibbs Politz, and Shriram Krishnamurthi.
\newblock Slimming languages by reducing sugar: A case for semantics-altering
  transformations.
\newblock In {\em 2015 {{ACM International Symposium}} on {{New Ideas}}, {{New
  Paradigms}}, and {{Reflections}} on {{Programming}} and {{Software}}
  ({{Onward}}!)}, pages 90--106, {Pittsburgh PA USA}, October 2015. {ACM}.

\bibitem{pombrioInferringTypeRules2018a}
Justin Pombrio and Shriram Krishnamurthi.
\newblock Inferring type rules for syntactic sugar.
\newblock In {\em Proceedings of the 39th {{ACM SIGPLAN Conference}} on
  {{Programming Language Design}} and {{Implementation}}}, pages 812--825,
  {Philadelphia PA USA}, June 2018. {ACM}.

\bibitem{stroustrupEvolvingLanguageReala}
Bjarne Stroustrup.
\newblock Evolving a language in and for the real world: C++ 1991-2006.
\newblock page~59, 2007.

\bibitem{hudakHistoryHaskellBeing2007a}
Paul Hudak, John Hughes, Simon Peyton~Jones, and Philip Wadler.
\newblock A history of {{Haskell}}: Being lazy with class.
\newblock In {\em Proceedings of the Third {{ACM SIGPLAN}} Conference on
  {{History}} of Programming Languages}, {San Diego California}, June 2007.
  {ACM}.

\bibitem{liskov1993history}
Barbara Liskov.
\newblock A history of clu.
\newblock {\em ACM SIGPLAN Notices}, 28(3):133--147, 1993.

\bibitem{smith1984reflection}
Brian~Cantwell Smith.
\newblock Reflection and semantics in lisp.
\newblock In {\em Proceedings of the 11th ACM SIGACT-SIGPLAN symposium on
  Principles of programming languages}, pages 23--35, 1984.

\bibitem{molerHistoryMATLAB}
Cleve Moler and Jack Little.
\newblock A {{History}} of {{MATLAB}}.
\newblock 4:67, 2020.

\bibitem{jagannathanMetalevelBuildingBlocks1994}
Suresh Jagannathan.
\newblock Metalevel building blocks for modular systems.
\newblock {\em ACM Transactions on Programming Languages and Systems},
  16(3):456--492, May 1994.

\bibitem{satoModuleGenerationRegret2020}
Yuhi Sato, Yukiyoshi Kameyama, and Takahisa Watanabe.
\newblock Module generation without regret.
\newblock In {\em Proceedings of the 2020 {{ACM SIGPLAN Workshop}} on {{Partial
  Evaluation}} and {{Program Manipulation}} - {{PEPM}} 2020}, pages 1--13, {New
  Orleans, LA, USA}, 2020. {ACM Press}.

\bibitem{spacekInheritanceSystemStructurala}
Petr Spacek, Christophe Dony, Chouki Tibermacine, and Luc Fabresse.
\newblock An inheritance system for structural \&\#38; behavioral reuse in
  component-based software programming.
\newblock page~10, 2012.

\bibitem{kay1993early}
Alan~C Kay.
\newblock The early history of smalltalk, hopl-ii: The second acm sigplan
  conference on history of programming languages, 1993.

\bibitem{craryModulesAbstractionParametric2017a}
Karl Crary.
\newblock Modules, abstraction, and parametric polymorphism.
\newblock In {\em Proceedings of the 44th {{ACM SIGPLAN Symposium}} on
  {{Principles}} of {{Programming Languages}}}, pages 100--113, {Paris France},
  January 2017. {ACM}.

\bibitem{stricklandContractsFirstclassModules}
T~Stephen Strickland and Matthias Felleisen.
\newblock Contracts for first-class modules.
\newblock page~12, 2009.

\bibitem{zingaro2007modern}
Daniel Zingaro.
\newblock Modern extensible languages.
\newblock {\em SQRL Report}, 47:16, 2007.

\bibitem{erdwegFrameworkExtensibleLanguages2013}
Sebastian Erdweg and Felix Rieger.
\newblock A framework for extensible languages.
\newblock In {\em Proceedings of the 12th International Conference on
  {{Generative}} Programming: Concepts \& Experiences - {{GPCE}} '13}, pages
  3--12, {Indianapolis, Indiana, USA}, 2013. {ACM Press}.

\bibitem{wadler2012}
Philip Wadler.
\newblock The expressibiltiy problem, 2012.

\bibitem{findler1998modular}
Robert~Bruce Findler and Matthew Flatt.
\newblock Modular object-oriented programming with units and mixins.
\newblock {\em ACM SIGPLAN Notices}, 34(1):94--104, 1998.

\bibitem{tennent1981principles}
Robert~D Tennent.
\newblock {\em Principles of programming languages}.
\newblock Prentice Hall PTR, 1981.

\bibitem{sestoft2017programming}
Peter Sestoft.
\newblock {\em Programming language concepts}.
\newblock Springer, 2017.

\bibitem{kitchenham2007guidelines}
Barbara Kitchenham and Stuart Charters.
\newblock Guidelines for performing systematic literature reviews in software
  engineering.
\newblock 2007.

\bibitem{irons1970experience}
Edgar~T Irons.
\newblock Experience with an extensible language.
\newblock {\em Communications of the ACM}, 13(1):31--40, 1970.

\bibitem{wiedenbeck1999comparison}
Susan Wiedenbeck, Vennila Ramalingam, Suseela Sarasamma, and Cynthia~L
  Corritore.
\newblock A comparison of the comprehension of object-oriented and procedural
  programs by novice programmers.
\newblock {\em Interacting with Computers}, 11(3):255--282, 1999.

\bibitem{sobraltextordmasculine2021old}
S{\'o}nia~Rolland Sobral{\textordmasculine}.
\newblock The old question: which programming language should we choose to
  teach to program?
\newblock 2021.

\bibitem{winograd1979beyond}
Terry Winograd.
\newblock Beyond programming languages.
\newblock {\em Communications of the ACM}, 22(7):391--401, 1979.

\bibitem{shrestha2020here}
Nischal Shrestha, Colton Botta, Titus Barik, and Chris Parnin.
\newblock Here we go again: Why is it difficult for developers to learn another
  programming language?
\newblock In {\em 2020 IEEE/ACM 42nd International Conference on Software
  Engineering (ICSE)}, pages 691--701. IEEE, 2020.

\bibitem{kohlbecker1986hygienic}
Eugene Kohlbecker, Daniel~P Friedman, Matthias Felleisen, and Bruce Duba.
\newblock Hygienic macro expansion.
\newblock In {\em Proceedings of the 1986 ACM Conference on LISP and Functional
  Programming}, pages 151--161, 1986.

\bibitem{kaczmarczyk2010identifying}
Lisa~C Kaczmarczyk, Elizabeth~R Petrick, J~Philip East, and Geoffrey~L Herman.
\newblock Identifying student misconceptions of programming.
\newblock In {\em Proceedings of the 41st ACM technical symposium on Computer
  science education}, pages 107--111, 2010.

\bibitem{piteira2013learning}
Martinha Piteira and Carlos Costa.
\newblock Learning computer programming: study of difficulties in learning
  programming.
\newblock In {\em Proceedings of the 2013 International Conference on
  Information Systems and Design of Communication}, pages 75--80, 2013.

\bibitem{tan2009learning}
Phit-Huan Tan, Choo-Yee Ting, and Siew-Woei Ling.
\newblock Learning difficulties in programming courses: undergraduates'
  perspective and perception.
\newblock In {\em 2009 International Conference on Computer Technology and
  Development}, volume~1, pages 42--46. IEEE, 2009.

\bibitem{wiedenbeck1999novice}
Susan Wiedenbeck and Vennila Ramalingam.
\newblock Novice comprehension of small programs written in the procedural and
  object-oriented styles.
\newblock {\em International Journal of Human-Computer Studies}, 51(1):71--87,
  1999.

\bibitem{lenarcic2006antiusability}
John Lenarcic.
\newblock The antiusability manifesto.
\newblock In {\em Proceedings of the 18th Australia conference on
  Computer-Human Interaction: Design: Activities, Artefacts and Environments},
  pages 337--339, 2006.

\bibitem{yeomans2019transformative}
Lucy Yeomans, Steffen Zschaler, and Kelly Coate.
\newblock Transformative and troublesome? students' and professional
  programmers' perspectives on difficult concepts in programming.
\newblock {\em ACM Transactions on Computing Education (TOCE)}, 19(3):1--27,
  2019.

\bibitem{charmaz1966search}
Kathy Charmaz and Jonathan~A Smith.
\newblock The search for meaning--grounded theory.
\newblock {\em Rethinking Methods in Psychology}, pages 27--49, 1966.

\bibitem{dershowitz1991rewrite}
Nachum Dershowitz, St{\'e}phane Kaplan, and David~A Plaisted.
\newblock Rewrite, rewrite, rewrite, rewrite, rewrite,….
\newblock {\em Theoretical Computer Science}, 83(1):71--96, 1991.

\bibitem{williamsFlexibleNotationSyntactic1982a}
M.~Howard Williams.
\newblock A {{Flexible Notation}} for {{Syntactic Definitions}}.
\newblock {\em ACM Transactions on Programming Languages and Systems},
  4(1):113--119, January 1982.

\bibitem{holt1979model}
Richard~C Holt and David~B Wortman.
\newblock A model for implementing euclid modules and type templates.
\newblock In {\em Proceedings of the 1979 SIGPLAN symposium on Compiler
  construction}, pages 8--12, 1979.

\bibitem{laddadFluidQuotesMetaprogramming2020}
Shadaj Laddad and Koushik Sen.
\newblock Fluid quotes: Metaprogramming across abstraction boundaries with
  dependent types.
\newblock In {\em Proceedings of the 19th {{ACM SIGPLAN International
  Conference}} on {{Generative Programming}}: Concepts and {{Experiences}}},
  pages 98--110, {Virtual USA}, November 2020. {ACM}.

\bibitem{krishnamurthiDesugaringPracticeOpportunities2015}
Shriram Krishnamurthi.
\newblock Desugaring in {{Practice}}: Opportunities and {{Challenges}}.
\newblock In {\em Proceedings of the 2015 {{Workshop}} on {{Partial
  Evaluation}} and {{Program Manipulation}}}, pages 1--2, {Mumbai India},
  January 2015. {ACM}.

\bibitem{huangMorphingStructurallyShaping2011}
Shan~Shan Huang and Yannis Smaragdakis.
\newblock Morphing: Structurally shaping a class by reflecting on others.
\newblock {\em ACM Transactions on Programming Languages and Systems},
  33(2):1--44, January 2011.

\bibitem{kohlbecker1987macro}
Eugene~E Kohlbecker and Mitchell Wand.
\newblock Macro-by-example: Deriving syntactic transformations from their
  specifications.
\newblock In {\em Proceedings of the 14th ACM SIGACT-SIGPLAN symposium on
  Principles of programming languages}, pages 77--84, 1987.

\bibitem{disneySweetenYourJavaScript2014}
Tim Disney, Nathan Faubion, David Herman, and Cormac Flanagan.
\newblock Sweeten your {{JavaScript}}: Hygienic macros for {{ES5}}.
\newblock In {\em Proceedings of the 10th {{ACM Symposium}} on {{Dynamic}}
  Languages - {{DLS}} '14}, pages 35--44, {Portland, Oregon, USA}, 2014. {ACM
  Press}.

\bibitem{ballantyneMacrosDomainspecificLanguages2020a}
Michael Ballantyne, Alexis King, and Matthias Felleisen.
\newblock Macros for domain-specific languages.
\newblock {\em Proceedings of the ACM on Programming Languages},
  4(OOPSLA):1--29, November 2020.

\bibitem{bakerMayaMultipleDispatchSyntaxa}
Jason Baker and Wilson~C Hsieh.
\newblock Maya: Multiple-{{Dispatch Syntax Extension}} in {{Java}}.
\newblock page~12.

\bibitem{tobinhochstadtLanguagesLibraries}
Sam {Tobin-Hochstadt}, Vincent {St-Amour}, Ryan Culpepper, Matthew Flatt, and
  Matthias Felleisen.
\newblock Languages as libraries.
\newblock page~10.

\bibitem{kiselyovMacrosThatCompose2002a}
Oleg Kiselyov.
\newblock Macros {{That Compose}}: Systematic {{Macro Programming}}.
\newblock In Gerhard Goos, Juris Hartmanis, Jan {van Leeuwen}, Don Batory,
  Charles Consel, and Walid Taha, editors, {\em Generative {{Programming}} and
  {{Component Engineering}}}, volume 2487, pages 202--217. {Springer Berlin
  Heidelberg}, {Berlin, Heidelberg}, 2002.

\bibitem{waddellExtendingScopeSyntactic1999a}
Oscar Waddell and R.~Kent Dybvig.
\newblock Extending the scope of syntactic abstraction.
\newblock In {\em Proceedings of the 26th {{ACM SIGPLAN}}-{{SIGACT}} Symposium
  on {{Principles}} of Programming Languages - {{POPL}} '99}, pages 203--215,
  {San Antonio, Texas, United States}, 1999. {ACM Press}.

\bibitem{culpepperSyntacticAbstractionComponent2005}
Ryan Culpepper, Scott Owens, and Matthew Flatt.
\newblock Syntactic {{Abstraction}} in {{Component Interfaces}}.
\newblock In David Hutchison, Takeo Kanade, Josef Kittler, Jon~M. Kleinberg,
  Friedemann Mattern, John~C. Mitchell, Moni Naor, Oscar Nierstrasz,
  C.~Pandu~Rangan, Bernhard Steffen, Madhu Sudan, Demetri Terzopoulos, Dough
  Tygar, Moshe~Y. Vardi, Gerhard Weikum, Robert Gl{\"u}ck, and Michael Lowry,
  editors, {\em Generative {{Programming}} and {{Component Engineering}}},
  volume 3676, pages 373--388. {Springer Berlin Heidelberg}, {Berlin,
  Heidelberg}, 2005.

\bibitem{zhangCompositionalProgramming2021}
Weixin Zhang, Yaozhu Sun, and Bruno C. D.~S. Oliveira.
\newblock Compositional {{Programming}}.
\newblock {\em ACM Transactions on Programming Languages and Systems},
  43(3):1--61, September 2021.

\bibitem{steele1996evolution}
Guy~L Steele and Richard~P Gabriel.
\newblock The evolution of lisp.
\newblock {\em History of programming languages---II}, pages 233--330, 1996.

\bibitem{dos2006specifying}
Gabriel Dos~Reis and Bjarne Stroustrup.
\newblock Specifying c++ concepts.
\newblock {\em ACM SIGPLAN Notices}, 41(1):295--308, 2006.

\bibitem{suttonDesignConceptLibraries2012}
Andrew Sutton and Bjarne Stroustrup.
\newblock Design of {{Concept Libraries}} for {{C}}++.
\newblock In David Hutchison, Takeo Kanade, Josef Kittler, Jon~M. Kleinberg,
  Friedemann Mattern, John~C. Mitchell, Moni Naor, Oscar Nierstrasz,
  C.~Pandu~Rangan, Bernhard Steffen, Madhu Sudan, Demetri Terzopoulos, Doug
  Tygar, Moshe~Y. Vardi, Gerhard Weikum, Anthony Sloane, and Uwe A{\ss}mann,
  editors, {\em Software {{Language Engineering}}}, volume 6940, pages 97--118.
  {Springer Berlin Heidelberg}, {Berlin, Heidelberg}, 2012.

\bibitem{stroustrupThrivingCrowdedChanging2020a}
Bjarne Stroustrup.
\newblock Thriving in a crowded and changing world: C++ 2006\textendash 2020.
\newblock {\em Proceedings of the ACM on Programming Languages},
  4(HOPL):1--168, June 2020.

\bibitem{macqueenUsingDependentTypesa}
David MacQueen.
\newblock Using {{Dependent Types}} to {{Express Modu}}|ar
  {{Struetr}}\textasciitilde re.
\newblock page~10.

\bibitem{symeEarlyHistory2020a}
Don Syme.
\newblock The early history of {{F}}\#.
\newblock {\em Proceedings of the ACM on Programming Languages}, 4(HOPL):1--58,
  June 2020.

\bibitem{hickeyHistoryClojure2020}
Rich Hickey.
\newblock A history of {{Clojure}}.
\newblock {\em Proceedings of the ACM on Programming Languages}, 4(HOPL):1--46,
  June 2020.

\bibitem{hansen1996monitors}
Per~Brinch Hansen.
\newblock Monitors and concurrent pascal: a personal history.
\newblock {\em History of programming languages---II}, pages 121--172, 1996.

\bibitem{jones1996using}
Mark~P Jones.
\newblock Using parameterized signatures to express modular structure.
\newblock In {\em Proceedings of the 23rd ACM SIGPLAN-SIGACT symposium on
  Principles of programming languages}, pages 68--78, 1996.

\bibitem{blume1999hierarchical}
Matthias Blume and Andrew~W Appel.
\newblock Hierarchical modularity.
\newblock {\em ACM Transactions on Programming Languages and Systems (TOPLAS)},
  21(4):813--847, 1999.

\bibitem{garyWhatRecursiveModulea}
Karl Gary, Robert Harper, and Sidd Puri.
\newblock What is a {{Recursive Module}}?
\newblock page~14.

\bibitem{wirthModula2Oberon2007a}
Niklaus Wirth.
\newblock Modula-2 and {{Oberon}}.
\newblock In {\em Proceedings of the Third {{ACM SIGPLAN}} Conference on
  {{History}} of Programming Languages}, {San Diego California}, June 2007.
  {ACM}.

\bibitem{stroustrup1996history}
Bjarne Stroustrup.
\newblock A history of c++ 1979--1991.
\newblock In {\em History of programming languages---II}, pages 699--769. 1996.

\bibitem{dreyerTypeSystemHigherOrdera}
Derek Dreyer, Karl Crary, and Robert Harper.
\newblock A {{Type System}} for {{Higher}}-{{Order Modules}}.
\newblock page~14.

\bibitem{kilpatrickBackpackRetrofittingHaskell2014}
Scott Kilpatrick, Derek Dreyer, Simon Peyton~Jones, and Simon Marlow.
\newblock Backpack: Retrofitting {{Haskell}} with interfaces.
\newblock In {\em Proceedings of the 41st {{ACM SIGPLAN}}-{{SIGACT Symposium}}
  on {{Principles}} of {{Programming Languages}}}, pages 19--31, {San Diego
  California USA}, January 2014. {ACM}.

\bibitem{rather1996evolution}
Elizabeth~D Rather, Donald~R Colburn, and Charles~H Moore.
\newblock The evolution of forth.
\newblock In {\em History of programming languages---II}, pages 625--670. 1996.

\bibitem{brookes2014essence}
Stephen Brookes, Peter~W O'hearn, and Uday Reddy.
\newblock The essence of reynolds, 2014.

\bibitem{tofte1994principal}
Mads Tofte.
\newblock Principal signatures for higher-order program modules.
\newblock {\em Journal of Functional Programming}, 4(3):285--335, 1994.

\bibitem{watanabeProgramGenerationML2017}
Takahisa Watanabe and Yukiyoshi Kameyama.
\newblock Program generation for {{ML}} modules (short paper).
\newblock In {\em Proceedings of the {{ACM SIGPLAN Workshop}} on {{Partial
  Evaluation}} and {{Program Manipulation}}}, pages 60--66, {Los Angeles CA
  USA}, December 2017. {ACM}.

\bibitem{morrisonAdHocApproach}
R~Morrison, A~Dearle, R~C~H Connor, and A~L Brown.
\newblock An ad hoc approach to the implementation of polymorphism.
\newblock page~30.

\bibitem{meyer1986type}
Albert~R Meyer and Mark~B Reinhold.
\newblock " type" is not a type.
\newblock In {\em Proceedings of the 13th ACM SIGACT-SIGPLAN symposium on
  Principles of programming languages}, pages 287--295, 1986.

\bibitem{wadlerHowMakeAdhoc1989a}
P.~Wadler and S.~Blott.
\newblock How to make ad-hoc polymorphism less ad hoc.
\newblock In {\em Proceedings of the 16th {{ACM SIGPLAN}}-{{SIGACT}} Symposium
  on {{Principles}} of Programming Languages - {{POPL}} '89}, pages 60--76,
  {Austin, Texas, United States}, 1989. {ACM Press}.

\bibitem{siekEssentialLanguageSupporta}
Jeremy Siek and Andrew Lumsdaine.
\newblock Essential {{Language Support}} for {{Generic Programming}}.
\newblock page~12.

\bibitem{kaminaLightweightScalableComponents2007}
Tetsuo Kamina and Tetsuo Tamai.
\newblock Lightweight scalable components.
\newblock In {\em Proceedings of the 6th International Conference on
  {{Generative}} Programming and Component Engineering - {{GPCE}} '07}, page
  145, {Salzburg, Austria}, 2007. {ACM Press}.

\bibitem{jansson1997polyp}
Patrik Jansson and Johan Jeuring.
\newblock Polyp—a polytypic programming language extension.
\newblock In {\em Proceedings of the 24th ACM SIGPLAN-SIGACT symposium on
  Principles of Programming Languages}, pages 470--482, 1997.

\bibitem{wolfingerAddingGenericityPlugin}
Reinhard Wolfinger, Markus L{\"o}berbauer, Markus Jahn, and Hanspeter
  M{\"o}ssenb{\"o}ck.
\newblock Adding genericity to a plug-in framework.
\newblock page~10.

\bibitem{wegnerUnificationDataProgram1983a}
Peter Wegner.
\newblock On the unification of data and program abstraction in {{Ada}}.
\newblock In {\em Proceedings of the 10th {{ACM SIGACT}}-{{SIGPLAN}} Symposium
  on {{Principles}} of Programming Languages - {{POPL}} '83}, pages 256--264,
  {Austin, Texas}, 1983. {ACM Press}.

\bibitem{teruelAccessControlReflection}
Camille Teruel, St{\'e}phane Ducasse, Damien Cassou, and Marcus Denker.
\newblock Access {{Control}} to {{Reflection}} with {{Object Ownership}}.
\newblock page~9.

\bibitem{huangExpressiveSafeStatic}
Shan~Shan Huang and Yannis Smaragdakis.
\newblock Expressive and {{Safe Static Reflection}} with {{MorphJ}}.
\newblock page~11.

\bibitem{politoBootstrappingInfrastructureBuild2015a}
Guillermo Polito, St{\'e}phane Ducasse, Noury Bouraqadi, and Luc Fabresse.
\newblock A bootstrapping infrastructure to build and extend {{Pharo}}-like
  languages.
\newblock In {\em 2015 {{ACM International Symposium}} on {{New Ideas}}, {{New
  Paradigms}}, and {{Reflections}} on {{Programming}} and {{Software}}
  ({{Onward}}!)}, pages 183--196, {Pittsburgh PA USA}, October 2015. {ACM}.

\bibitem{miaoCompiletimeReflectionMetaprogramming2014}
Weiyu Miao and Jeremy Siek.
\newblock Compile-time reflection and metaprogramming for {{Java}}.
\newblock In {\em Proceedings of the {{ACM SIGPLAN}} 2014 {{Workshop}} on
  {{Partial Evaluation}} and {{Program Manipulation}} - {{PEPM}} '14}, pages
  27--37, {San Diego, California, USA}, 2014. {ACM Press}.

\bibitem{wirfs-brockJavaScriptFirst202020a}
Allen {Wirfs-Brock} and Brendan Eich.
\newblock {{JavaScript}}: The first 20 years.
\newblock {\em Proceedings of the ACM on Programming Languages},
  4(HOPL):1--189, June 2020.

\bibitem{oderskySimplicitlyFoundationsApplications2018a}
Martin Odersky, Olivier Blanvillain, Fengyun Liu, Aggelos Biboudis, Heather
  Miller, and Sandro Stucki.
\newblock Simplicitly: Foundations and applications of implicit function types.
\newblock {\em Proceedings of the ACM on Programming Languages}, 2(POPL):1--29,
  January 2018.

\bibitem{wirth1996recollections}
Niklaus Wirth.
\newblock Recollections about the development of pascal.
\newblock In {\em History of programming languages---II}, pages 97--120. 1996.

\bibitem{servettoMetacircularLanguageActivea}
Marco Servetto and Elena Zucca.
\newblock A meta-circular language for active libraries.
\newblock page~10.

\bibitem{macqueenHistoryStandardML2020a}
David MacQueen, Robert Harper, and John Reppy.
\newblock The history of {{Standard ML}}.
\newblock {\em Proceedings of the ACM on Programming Languages},
  4(HOPL):1--100, June 2020.

\bibitem{lewisImplicitParametersDynamic2000a}
Jeffrey~R. Lewis, John Launchbury, Erik Meijer, and Mark~B. Shields.
\newblock Implicit parameters: Dynamic scoping with static types.
\newblock In {\em Proceedings of the 27th {{ACM SIGPLAN}}-{{SIGACT}} Symposium
  on {{Principles}} of Programming Languages - {{POPL}} '00}, pages 108--118,
  {Boston, MA, USA}, 2000. {ACM Press}.

\bibitem{homerFirstclassDynamicTypes2019a}
Michael Homer, Timothy Jones, and James Noble.
\newblock First-class dynamic types.
\newblock In {\em Proceedings of the 15th {{ACM SIGPLAN International
  Symposium}} on {{Dynamic Languages}}}, pages 1--14, {Athens Greece}, October
  2019. {ACM}.

\bibitem{gelernterEnvironmentsFirstClass1987a}
D.~Gelernter, S.~Jagannathan, and T.~London.
\newblock Environments as first class objects.
\newblock In {\em Proceedings of the 14th {{ACM SIGACT}}-{{SIGPLAN}} Symposium
  on {{Principles}} of Programming Languages - {{POPL}} '87}, pages 98--110,
  {Munich, West Germany}, 1987. {ACM Press}.

\bibitem{homerAPIsLanguagesGeneralising2015a}
Michael Homer, Timothy Jones, and James Noble.
\newblock From {{APIs}} to languages: Generalising method names.
\newblock In {\em Proceedings of the 11th {{Symposium}} on {{Dynamic
  Languages}}}, pages 1--12, {Pittsburgh PA USA}, October 2015. {ACM}.

\bibitem{scott2000programming}
Michael~Lee Scott.
\newblock {\em Programming language pragmatics}.
\newblock Morgan Kaufmann, 2000.

\bibitem{glazeAbstractingAbstractControl2014b}
Dionna Glaze and David Van~Horn.
\newblock Abstracting abstract control.
\newblock In {\em Proceedings of the 10th {{ACM Symposium}} on {{Dynamic}}
  Languages}, pages 11--22, {Portland Oregon USA}, October 2014. {ACM}.

\bibitem{altmanKeynoteLanguageOptimizer2011a}
Erik Altman.
\newblock Keynote {{I}}: The language, optimizer, and tools mess.
\newblock In {\em International {{Symposium}} on {{Code Generation}} and
  {{Optimization}} ({{CGO}} 2011)}, pages xxx--xxx, {Chamonix, France}, April
  2011. {IEEE}.

\bibitem{boehmThreadsCannotBea}
Hans-J Boehm.
\newblock Threads {{Cannot Be Implemented As}} a {{Library}}.
\newblock page~8.

\bibitem{parkinsonSeparationLogicAbstractiona}
Matthew Parkinson and Gavin Bierman.
\newblock Separation {{Logic}} and {{Abstraction}}.
\newblock page~12.

\bibitem{fournetFullyAbstractCompilationa}
Cedric Fournet, Nikhil Swamy, Juan Chen, Pierre-Evariste Dagand, Pierre-Yves
  Strub, and Benjamin Livshits.
\newblock Fully abstract compilation to {{JavaScript}}.
\newblock page~13.

\bibitem{naurPAPEREUROPEANSIDEa}
Peter Naur.
\newblock {{PAPER}}: {{THE EUROPEAN SIDE OF THE LAST PHASE OF THE DEVELOPMENT
  OF ALGOL}} 60.
\newblock page~48.

\bibitem{brightOriginsProgrammingLanguage2020}
Walter Bright, Andrei Alexandrescu, and Michael Parker.
\newblock Origins of the {{D}} programming language.
\newblock {\em Proceedings of the ACM on Programming Languages}, 4(HOPL):1--38,
  June 2020.

\bibitem{lamPuttingSemanticsSemantic2020a}
Patrick Lam, Jens Dietrich, and David~J. Pearce.
\newblock Putting the semantics into semantic versioning.
\newblock In {\em Proceedings of the 2020 {{ACM SIGPLAN International
  Symposium}} on {{New Ideas}}, {{New Paradigms}}, and {{Reflections}} on
  {{Programming}} and {{Software}}}, pages 157--179, {Virtual USA}, November
  2020. {ACM}.

\bibitem{chambersDebateLanguageTool2000a}
Craig Chambers, Bill Harrison, and John Vlissides.
\newblock A debate on language and tool support for design patterns.
\newblock In {\em Proceedings of the 27th {{ACM SIGPLAN}}-{{SIGACT}} Symposium
  on {{Principles}} of Programming Languages - {{POPL}} '00}, pages 277--289,
  {Boston, MA, USA}, 2000. {ACM Press}.

\end{thebibliography}

\newpage
\appendices
\section{Review traceability}
\label{app:review-traceability}
This appendix includes references to all relevant data points from the conducted review, i.e. titles of publications during the different stages of filtering. All data was collected using a custom (open-source) retrieval service\footnote{\url{https://github.com/sebastiannicolajsen/dblp-fetcher}} built on top of \emph{dblp.org}. This service outputs its data as a \emph{json} blob. All publications are thereby listed with both title, doi, and (potentially) comments as well as labels. Labels are used to indicate that publications have been selected for further processing. Do note that as a consequence of utilising \textit{dblp}, the data exported contains more than simply the needed publications, it also contains (i) titles of the proceedings, and (ii) publications from some other conferences (e.g. OOPSLA). The amount of entries may therefore not necessarily be used to (directly) deduce the filtered amount of publications.

\subsection{Unfiltered publications}
\label{app:review-traceability:unfiltered}

All unfiltered states can be found here: \\ \noindent
\url{https://drive.google.com/drive/folders/1Br28i-hg-Bx6wBS_ha1EDRFKVG79Yyiv?usp=sharing}

\subsection{Publications filtered based on title}
\label{app:review-traceability:filter1}

All publications filtered based on title can be found here: \\ \noindent
\url{https://drive.google.com/drive/folders/1u3UVcy5x9bKqHpmPo-3Oceq3cPRP76bq?usp=sharing}

\subsection{Publications filtered based on abstract and conclusion}
\label{app:review-traceability:filter2}
All publications filtered based on abstract and conclusion are combined into a single file, where excluded publications are also present (but without the \texttt{"label":true} attribute), and can be found here:\\ \noindent
\url{https://drive.google.com/file/d/17_P2zf11TCXNN7f1T-0PVnjqByd_jPGe/view?usp=sharing}

\subsection{Publication quality}
\label{app:review-traceability:filter3}
The quality assessment utilised the Zotero platform to manage articles and their quality. The below file provides the Zotero library. Here, the \texttt{None} label was used to indicate no relevance. However, if an article also contains the label \texttt{Almost none} it indicates a partial relevance.
\\
\url{https://drive.google.com/file/d/1PWPxcy0XZkoffjmsp_FIq7Rc1EH0Icd_/view?usp=sharing}

\section{Keywords for selection criteria}
\label{app:selection-criteria-keywords}

\noindent\textbf{Self defined keywords (16)}
\noindent Extensibility, Extensible, Extension, Language definition, Language constructs, Language capabilities, Language style, Library, syntax, semantic, (type), Abstraction, parameter, polymorphism, reflection, generics
\\\\
\noindent\textbf{All keywords from preliminary investigation (132)}\\
\noindent programming languages,extensible, compilers bootstrapping, ambiguity, Object-oriented programming, Empirical studies of programming, Novice programmers, Learning to program, programming languages, Undergraduate studies, introduction to programming, programming, programming languages, programming systems, systems development, object-orientation, programming; novices, programming difficulties, software visualization, Human-centered computing, Empirical studies in HCI, Software and its engineering, Programming teams, Curriculum, Concept Inventory, Programming, Misconceptions, Pedagogy, CS1, Computer programming, novice programming, learning environments, programming, programming difficulties, learning, game-based learning interest, motivation, education game, analogies related to social aspects, pure and applied mathematics, language quality, language development, artificial auxiliary languages, literature, style, descriptive and prescriptive attitudes, User Interface Design Theory,  Social and professional topics, Software engineering education, Software and its engineering, Threshold concepts, learning programming, focus groups, computer science curriculum, accidental complexities, Continuous assessment, Formative feedback, Student motivation Portfolios, Learning to program, Teaching programming
\\\\
\noindent\textbf{Usable from preliminary investigation (duplicates removed, 65)}\\
\noindent programming languages, extensible, compilers bootstrapping, programming difficulties, ambiguity,  language quality, language development, style, Empirical studies of programming, Concept Inventory, Misconceptions, analogies related to social aspects,  Threshold concepts, accidental complexities,
\\\\
\noindent\textbf{Potential usable keywords from preliminary investigation}\\
\noindent Learning to program, Teaching programming, Software and its engineering, motivation, artificial auxiliary languages,

\section{Quality assessment questions}
\label{app:quality-assessment-questions}
The primary questions asked:

\begin{itemize}
    \item Does the article discuss extensibility? (Directly or indirectly)
    \item Does the article discuss concepts of extensibility?
\end{itemize}

Secondary questions asked:

\begin{itemize}
    \item Does the article discuss a specific concept of extensibility? (In a single language? Across languages?)
    \item Does the article discuss a specific language and extensibility? (A single element of extensibility? A single element of extensibility?)
\end{itemize}

\end{document}